\documentclass[reprint,superscriptaddress,amsmath,amssymb,aps,prb,floatfix]{revtex4-2}
\usepackage{graphicx}
\usepackage{color}

\usepackage{dcolumn}
\usepackage{upgreek}
\usepackage{gensymb}
\usepackage{float}
\usepackage{bm}
\usepackage{xcolor}
\usepackage{physics}
\usepackage{longtable}
\usepackage{comment}
\usepackage{graphicx}
\usepackage{subfigure}
\usepackage{mhchem}

\usepackage{xr}
\makeatletter
\newcommand*{\addFileDependency}[1]{
	\typeout{(#1)}
	\@addtofilelist{#1}
	\IfFileExists{#1}{}{\typeout{No file #1.}}
}
\makeatother

\newcommand*{\myexternaldocument}[1]{
	\externaldocument{#1}
	\addFileDependency{#1.tex}
	\addFileDependency{#1.aux}
}

\myexternaldocument{supps_052926}

\begin{document}
	
\title{Phase control of magnon-phonon coupling via magnetic field}

\author{Yasuhiro Todaka}
\affiliation{Department of Physics, The University of Tokyo, Hongo, Tokyo 113-0033, Japan}

\author{Motoki Asano}
\affiliation{Basic Research Laboratories, NTT, Inc., Atsugi, Kanagawa 243-0198, Japan}

\author{Isamu Yasuda}
\affiliation{Department of Physics, The University of Tokyo, Hongo, Tokyo 113-0033, Japan}

\author{Masashi Kawaguchi}
\affiliation{Department of Physics, The University of Tokyo, Hongo, Tokyo 113-0033, Japan}

\author{Daiki Hatanaka}
\affiliation{Basic Research Laboratories, NTT, Inc., Atsugi, Kanagawa 243-0198, Japan}

\author{Masamitsu Hayashi}
\affiliation{Department of Physics, The University of Tokyo, Hongo, Tokyo 113-0033, Japan}
\affiliation{Trans-scale quantum science institute (TSQS), The University of Tokyo, Hongo, Tokyo 113-0033, Japan}

\date{\today}

\begin{abstract}
We study the phase of the coupling between magnons and surface acoustic wave (SAW) phonons in magnetic thin films. 
The coupling constant changes from a real to a complex number as the external magnetic field is reduced.
Below a transition field, the imaginary coupling constant allows SAW phonons to couple to overdamped magnons whose resonance frequency is close to zero and far from the SAW resonance.
The strength of the imaginary coupling constant and the magnitude of the transition field both scale with magnetic damping.
We find the coupling produces a broad, pronounced minimum in the SAW transmittance spectrum near zero magnetic field in a Ni/Ru/Ni synthetic antiferromagnet with large magnetic damping. 
These results demonstrate that the phase of the complex magnon-phonon coupling constant can be tuned via magnetic field in strongly damped magnets, offering a platform to explore novel regimes of magnon-phonon interactions.
\end{abstract} 

\maketitle

\section{INTRODUCTION}
The coupling of magnons with other quanta in solid state systems is attracting significant interest recently as potential applications in hybrid quantum systems and/or magnonic devices are being considered\cite{awschalom2021ieee,pirro2021nrevmat,yuan2022physrep}.
Studies have shown that magnons can couple to superconducting qubits\cite{tabuchi2015science}, photons\cite{zhang2014prl,tabuchi2014prl,xzhang2016sciadv,hou2019prl},  phonons\cite{holanda2018nphys,berk2019ncomm,an2020prb,hioki2022commphys,hatanaka2022prap,hwang2024prl,matsumoto2024nanolett} and magnons of different modes\cite{li2020prl,shiota2020prl,kamimaki2020prap,wang2024ncomm}.
Significant effort has been put on exploring means and finding materials to increase the coupling strength, aiming to reach the ultrastrong coupling regime as well as to understand the underlying physics of the coupling mechanism.
For the latter, it has been reported that the coupling constant can be a complex number with its real and imaginary parts playing a different role\cite{xu2016nature,kohler2018prl,harder2018prl,carrara2024prl}.
The real part is commonly known as the coherent coupling constant, and causes the anticrossing of different quanta.
The imaginary part, in contrast, has been reported to induce synchronization of two quanta under certain conditions\cite{harder2018prl,grigoryan2019prb,lu2023jap}.

Here we demonstrate that the phase of the complex magnon-phonon coupling constant in ferromagnetic thin films can be tuned simply by varying the magnetic field. 
Specifically, we investigate the interaction between acoustic magnons in a synthetic antiferromagnet (SAF) and coherent phonons generated by surface acoustic waves (SAWs)\cite{weiler2011prl,dreher2012prb,thevenard2014prb,gowtham2015jap}.
SAF is chosen since it is easier to experimentally probe the imaginary coupling constant.
At a transition magnetic field, corresponding to an exceptional point for special material parameters, the coupling constant changes from a real to a complex number\cite{xu2016nature,partanen2019prb,liu2019sciadv,tserkovnyak2020prr}.
The emergence of an imaginary component below the transition field enables SAW phonons to couple with overdamped magnons whose resonance frequency lies far from that of the phonons, offering an alternative mechanism for magnon-phonon coupling.
Because both the transition field and the imaginary part of the coupling constant scale with the magnetic damping constant, materials with high magnetic damping are particularly well-suited for exploring the complex coupling behavior.

\section{MODEL CALCULATIONS}
We start from the equation of motion of the magnetization (i.e., the Landau-Lifshitz-Gilbert (LLG) equation) and the elastic wave equation, which can be reduced into the following form:
\begin{equation}
\begin{aligned}
\dot{a} =& \large[ i (\omega - \omega_a) - \frac{\kappa_a}{2} \large] a - i g_{am} m,\\
\dot{m} =& \large[ i (\omega - \omega_m) - \frac{\kappa_m}{2} \large] m - i g_{ma} a,
	\label{eq:eom:coupled}
\end{aligned}
\end{equation}
where $a$ and $m$ are the complex amplitudes of the phonons and magnons, respectively. These amplitudes are regarded as the annihilation operators via the second quantization. 
$\omega$ is the frequency of the driving force, $\omega_a$ ($\omega_m$) and $\kappa_a/2$ $(\kappa_m/2)$ are the eigenfrequency and relaxation rate of the phonons (magnons), and $g_{am} = |g_{am}| e^{i \phi_{am}}$ and $g_{ma} = |g_{ma}| e^{i \phi_{ma}}$ represent the coupling constants.
These equations derive from the following effective Hamiltonian:
\begin{align}
    \mathcal{H}=\hbar\omega_aa^\dagger a+\hbar\omega_m m^\dagger m+\hbar g(a^\dagger m+e^{i\Phi}m^\dagger a),
	\label{eq:H}
\end{align}
where $g$ and $\Phi$ are the magnitude and phase of the general coupling constant which satisfy $g = \sqrt{|g_{am}||g_{ma}|}$ and $\Phi = \phi_{am}+\phi_{ma}$.
Previous studies have primarily focused on the energy-conserving coherent magnon-phonon coupling, i.e., coupling with $\Phi=0$\cite{hatanaka2022prap,hwang2024prl,matsumoto2024nanolett}. In general, however, the coupling includes a non-energy-conserving part with nonzero $\Phi$.
The special case of $\Phi = \pi$ is often referred to as the "dissipative coupling"\cite{xu2016nature,kohler2018prl,harder2018prl,grigoryan2019prb,lu2023jap}.

We consider a symmetric SAF where the magnetic properties of the two ferromagnetic layers are the same: that is, for both layers, the saturation magnetization is $M_\mathrm{s}$, the magnetic damping constant is $\alpha$, the gyromagnetic ratio is $\gamma$ and the thickness is $t_{\mathrm{F}}$.
The interlayer exchange coupling field is defined as $H_\mathrm{E}$.
To simplify discussion, we first neglect the magnetic anisotropy and the magnetic dipole interaction: their influence will be discussed later.
An external in-plane magnetic field $\mu_0 \bm{H} = \mu_0 H(\cos\varphi_H, \sin\varphi_H, 0)$ is applied to the SAF.
$\mu_0$ is the magnetic permeability, $H$ and $\varphi_H$ are the magnitude and angle (with respect to the $x$ axis) of the magnetic field.
The $x$ axis is defined as the direction along which the SAW propagates [see Fig.~\ref{fig:setup:spectra}(a)].
From the LLG equation, $\omega_m$ is expressed as
\begin{equation}
\begin{gathered}
\omega_m=\mathrm{Im}\left[\dfrac{\mu_0\gamma}{2(1+\alpha^2)}\sqrt{\lambda}\right].
	\label{eq:omega}
\end{gathered}
\end{equation}
$\lambda$ for the symmetric SAF is given as
\begin{equation}
\begin{gathered}
\lambda\equiv-8H_{\mathrm{E}}(2H_{\mathrm{E}}+M_{\mathrm{s}})\cos^2{\varphi_0} +\alpha^2(2H_{\mathrm{E}}\sin^2{\varphi_0}+M_{\mathrm{s}})^2,
	\label{eq:lambda:SAF}
\end{gathered}
\end{equation}
where $2 \varphi_0$ is the relative angle of the magnetization between the two ferromagnetic layers\cite{asano2023prb}:
\begin{equation}
\begin{aligned}
\varphi_0 &= 
\begin{cases}
\cos^{-1}\left(\dfrac{H}{2H_{\mathrm{E}}}\right) & \text{if } \left|H\right| \le 2H_{\mathrm{E}}, \\
0 & \text{otherwise.}
\end{cases}
\label{eq:phi0}
\end{aligned}
\end{equation}
Equation~(\ref{eq:omega}) indicates that $\omega_m = 0$ when $\lambda>0$.
Under such circumstance, there is no magnon eigenmode and the magnetic system operates in the so-called fully damped regime. 
When $\lambda<0$, in contrast, a well-defined eigenmode appears and the system enters the underdamped regime.
$\lambda=0$ is the transition point where the LLG equation possesses a degenerate solution.
The state is equivalent to the "exceptional point" in the magnon system\cite{tserkovnyak2020prr}.

From the coupled equations of motion [Eq.~(\ref{eq:eom:coupled})], an analytical form for $g$ and $\Phi$ can be derived, which reads (see also the Appendix)
\begin{equation}
\begin{aligned}
	g e^{ i \frac{\Phi}{2} } =
	\begin{cases}
	\frac{A}{\sqrt{2}} \left(1 - i \right) \lambda^{-\frac{1}{4}} |\sin{2\varphi_H}\cos{2\varphi_0}|, \ \lambda > 0,\\
	A |\lambda|^{-\frac{1}{4}} |\sin{2\varphi_H}\cos{2\varphi_0}|, \ \lambda < 0,
	\end{cases}
	\label{eq:geff}
\end{aligned}
\end{equation}
where $A$ is a constant.
We have $\Phi=0$ for $\lambda<0$ and $\Phi=-\pi/2$ for $\lambda>0$.
The coupling constant therefore changes from a real number to a complex number as one moves from the underdamped to the fully damped regimes.
Previous studies have defined the "dissipative coupling" as a state with $\Phi = \pi$\cite{xu2016nature,kohler2018prl,harder2018prl,lu2023jap}.
Here we have $\Phi = - \pi/2$ since $\Phi = \phi_{am}+\phi_{ma}$, where $\phi_{am} = 0$ and $\phi_{ma} = - \pi/2$ in Eq.~(\ref{eq:eom:coupled}) for the fully damped mode.

\begin{figure}[bt]
	\centering
	\includegraphics[width=1\linewidth]{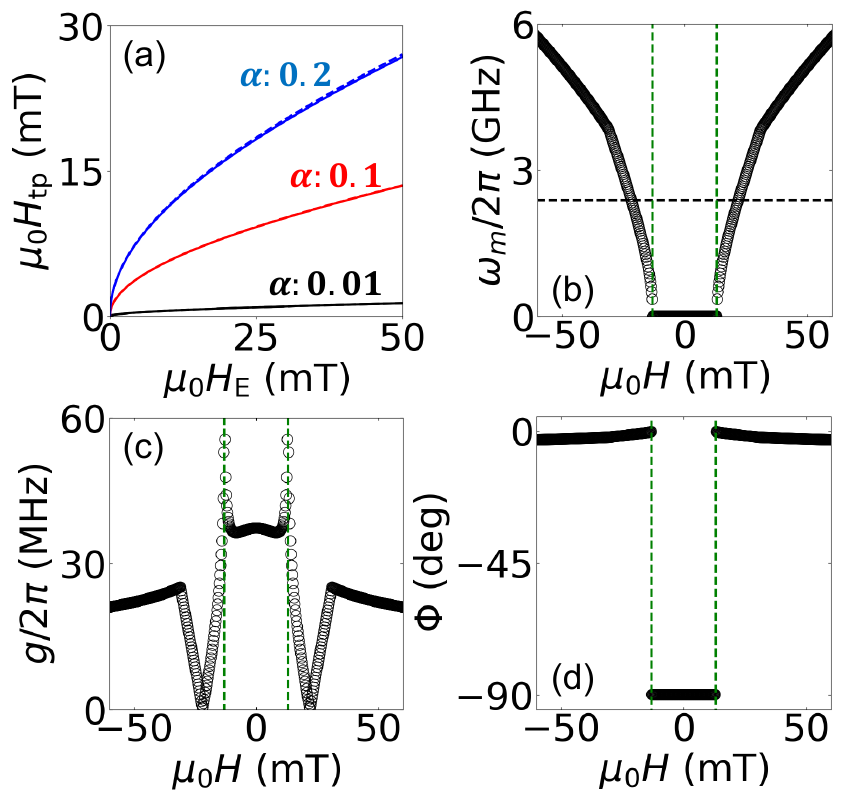}
	\caption{(a) $\mu_0 H_\mathrm{tp}$ [Eq.~(\ref{eq:hep})] plotted against the interlayer exchange coupling field $\mu_0 H_\mathrm{E}$. Solid and dashed lines show results from numerical calculations and Eq.~(\ref{eq:hep}), respectively. The magnetic damping constant $\alpha$ is varied: results are shown by different colors.  
	(b)-(d) Magnetic field $\mu_0 H$ ($\varphi_H = 135$ deg) dependence of the acoustic magnon eigenfrequency $\omega_m / 2 \pi $ (b), magnitude $g$ (c) and phase $\Phi$ (d) of the magnon-phonon coupling constant. The vertical dashed line indicates $|H| = H_\mathrm{tp}$. The horizontal dashed line in (b) shows the SAW resonance frequency used in the experiments. Parameters used in the calculations are listed in Table~\ref{table:par} of the Appendix. Here the uniaxial magnetic anisotropy field $\mu_0 H_\mathrm{u}$ and the dipolar interaction field $\mu_0 H_\mathrm{D}$ are set to zero.
    \label{fig:coupling}
	}
\end{figure}

The field range in which the $\Phi = - \pi/2$ coupling appears is given by $|H| \leq H_\mathrm{tp}$, where under $\alpha \ll 1$, the transition field $H_\mathrm{tp}$ is given by
\begin{equation}
\begin{aligned}
    H_\mathrm{tp} \approx \frac{\alpha}{2} \sqrt{ 2 H_{\mathrm{E} } ( 2 H_{\mathrm{E}} + M_{\mathrm{s}} ) }.
    \label{eq:hep}
\end{aligned}
\end{equation}
In Fig.~\ref{fig:coupling}(a), we plot $\mu_0H_\mathrm{tp}$ as a function of $\mu_0H_\mathrm{E}$, where the solid and dashed lines show results from numerical calculations and Eq.~(\ref{eq:hep}), respectively.
Colors indicate $\mu_0 H_\mathrm{tp}$ calculated with different $\alpha$.
Clearly, Eq.~(\ref{eq:hep}) provides a good estimate of $\mu_0H_\mathrm{tp}$ even when $\alpha$ is as large as 0.2.
As dictated by Eq.~(\ref{eq:hep}), $\mu_0 H_\mathrm{tp}$ increases with increasing $\alpha$ and $H_\mathrm{E}$.

We note that the $\Phi = - \pi/2$ coupling is not limited to a symmetric SAF and can be found in, for example, a single ferromagnetic layer film.
However, its field range is typically small.
The resonance frequency of a single layer film is given by Eq.~(\ref{eq:omega}) with 
\begin{equation}
\begin{gathered}
\lambda = -4 H (H + M_{\mathrm{s}} ) + \alpha^2 M_{\mathrm{s}}^2.
	\label{eq:lambda:single}
\end{gathered}
\end{equation}
Substituting the same material parameters, we estimate $\mu_0H_\mathrm{tp} = 0.016, 1.6, 6.3$ mT for the single layer film when $\alpha = 0.01, 0.1, 0.2$, respectively.
In all cases, $\mu_0 H_\mathrm{tp}$ is smaller than that of the SAF with $\mu_0 H_\mathrm{E} \gtrsim 10$ mT. 
Note that multidomain states often appear near zero field for the single layer films with small $H_{\mathrm{tp}}$, which can obscure the presence of the fully damped regime.
Thanks to its reduced stray field, the multidomain state is less likely to occur in the SAF.

Next, we present results from numerical calculations to show how the coupling constant varies with the magnetic field.
Figure~\ref{fig:coupling}(b) shows the $\mu_0 H$ dependence of the eigenfrequency $\omega_m / 2 \pi$ of the acoustic-mode magnons in the symmetric SAF. 
The angle $\varphi_H$ between the $x$ axis and the in-plane magnetic field is 135 deg.
The horizontal dashed line shows the SAW frequency, i.e., the SAW resonance frequency, which is set equal to that used in the experiments. 
The corresponding magnitude $g$ and phase $\Phi$ of the magnon-phonon coupling constant are displayed in Figs.~\ref{fig:coupling}(c) and \ref{fig:coupling}(d), respectively.
The vertical green dotted lines indicate $|H| = H_\mathrm{tp}$.
As is evident, $\omega_m / 2 \pi = 0$ and $\Phi = -\pi/2$ when $|H| \leq H_\mathrm{tp}$.
When $|H| \geq H_\mathrm{tp}$, $\omega_m / 2 \pi > 0$ and $\Phi$ drops to zero.
Interestingly, $g$ takes a nonzero value when $|H| \leq H_\mathrm{tp}$, indicating that SAW phonons can couple with the zero frequency (fully damped) magnons.
Note that $g$ drops to zero at $|H| = \sqrt{2} H_\mathrm{E}$ due to the $\cos 2 \varphi_0$ dependence of the coupling constant [see Eq.~(\ref{eq:geff})].

\begin{figure}[tb]
	\centering
	\includegraphics[width=1.0\linewidth]{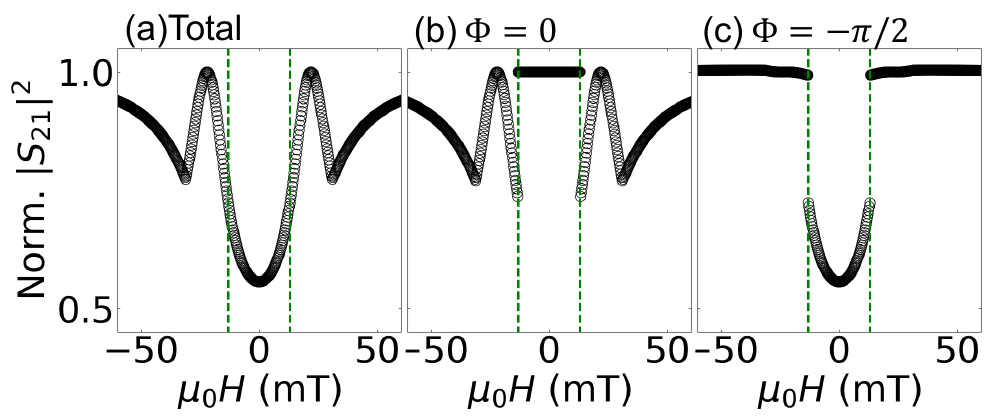}
	\caption{
    (a)-(c) Normalized SAW transmittance $|S_{21}|^2$ vs in-plane magnetic field $\mu_0 H$ ($\varphi_H = 135$ deg). $|S_{21}|^2$ is calculated assuming $\Phi = 0$ (b) and $\Phi = -\pi/2$ (c). The sum of the two is presented in (a). Parameters used in the calculations are listed in Table~\ref{table:par} of the Appendix. Here the uniaxial magnetic anisotropy field $\mu_0 H_\mathrm{u}$ and the dipolar interaction field $\mu_0 H_\mathrm{D}$ are set to zero.
	\label{fig:transmission}
	}
\end{figure}

The SAW transmittance $|S_{21}|^2$, which can be readily measured in experiments, is estimated by calculating the ratio of the elastic wave amplitudes that enter and traverse the SAF.
See the Appendix for the details of the calculations.
Figure~\ref{fig:transmission}(a) shows the $\mu_0 H$ dependence of the calculated $|S_{21}|^2$ normalized by its value at large $|\mu_0 H|$.
The results show that there are minimums at $\mu_0 H \sim \pm31$ mT and near zero field.
To identify the origin of the minimums, contributions from magnon modes with coupling phase of $\Phi = 0$ and $- \pi/2$ are shown in Figs.~\ref{fig:transmission}(b) and \ref{fig:transmission}(c).
Here we turn off one of the couplings and calculate $|S_{21}|^2$.
The minimum at $\mu_0 H \sim \pm31$ mT is caused by the $\Phi = 0$ coupling whereas the one at zero field is due to the $\Phi = - \pi/2$ coupling.
The former, often referred to as the coherent coupling, manifests itself in the $|S_{21}|^2$ spectrum when the SAW phonon and magnon eigenfrequencies are close to each other.
In contrast, the $\Phi = - \pi/2$ coupling does not require the magnon eigenfrequency to be close to that of the SAW phonon.

The calculations imply that SAW transmittance takes a minimum near zero field due to the emergence of $\Phi = - \pi/2$ coupling.
According to Eqs.~(\ref{eq:lambda:SAF}) and (\ref{eq:geff}), the magnitude of the $\Phi = - \pi/2$ coupling increases with the magnetic damping constant $\alpha$.
Moreover, Eq.~(\ref{eq:hep}) indicates that the field range in which the $\Phi = - \pi/2$ coupling occurs scales with $\alpha$.
We, therefore, use a symmetric SAF composed of Ni/Ru/Ni to explore the $\Phi = - \pi/2$ coupling since $\alpha$\cite{walowski2008jpd} and the magnetoelastic coupling\cite{bozorth1993book} of Ni are one of the largest among the $3d$ transition metals and their alloys.

\section{EXPERIMENTAL RESULTS}
\begin{figure}[tb]
	\centering
	\includegraphics[width=1\linewidth]{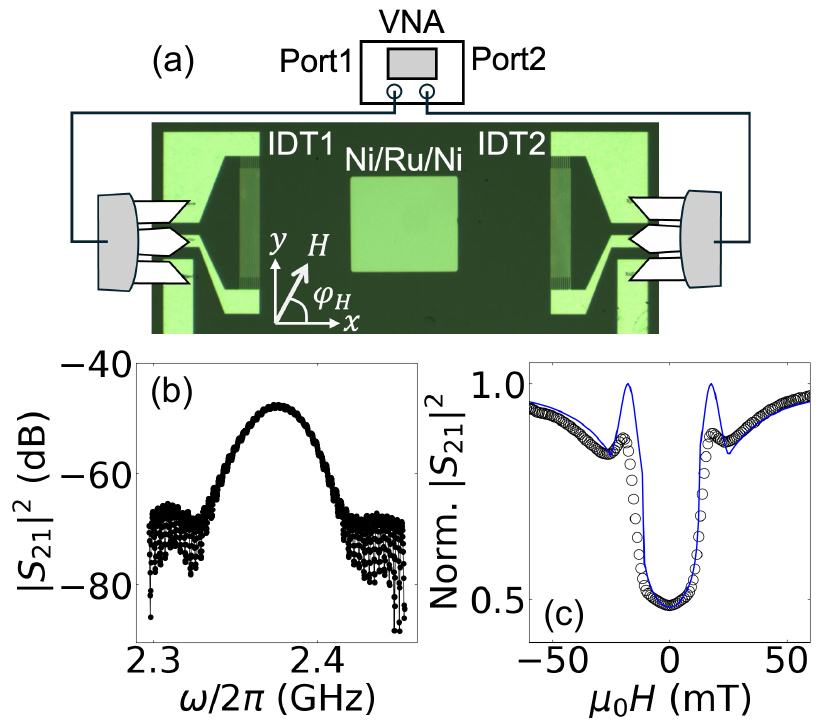}
	\caption{(a) Schematic illustration of the device used to study the SAW transmission across the Ni/Ru/Ni SAF. 
	(b) SAW transmittance $|S_{21}|^2$ plotted against the SAW excitation frequency $\omega / 2 \pi$. (c) Dependence of the normalized $|S_{21}|^2$ on the magnetic field $\mu_0 H$ ($\varphi_H = 135$ deg). 
	Black circles are the experimental results, the blue line shows the fitting result. Parameters used in the calculations are listed in Table~\ref{table:par} of the Appendix.
		\label{fig:setup:spectra}
	}
\end{figure}

All films are deposited on piezoelectric Y+128$^\mathrm{o}$-cut LiNbO$_{3}$ substrates using radio frequency (RF) magnetron sputtering.
The planar structures are formed using standard photolithography.
First, pairs of interdigital transducers (IDTs), made of Ta(5)/Al(50)/Pt(5) (thickness in units of nanometers), are patterned on the substrate. 
Subsequently, a rectangular element ($450~\mathrm{\mu m} \times 400~\mathrm{\mu m}$) comprised of a Ni/Ru/Ni SAF is formed in between the IDT pair. 
The complete film structure of the SAF is sub./Ta(1)/Pt(3)/Ni(5)/Ru($t_\mathrm{Ru}$)/Ni(5)/Ru(1)/MgO(2) /Ta(1). 
$t_\mathrm{Ru}$ is varied to control the interlayer exchange coupling between the two Ni layers.
Here we show representative results from $t_\mathrm{Ru} \sim 0.9$ nm.
Pt is included in the film stack to promote smooth growth of Ni as well as to increase the magnetic damping constant of the adjacent Ni layer\cite{mizukami2002prb,tserkovnyak2002prl,walowski2008jpd}.
An optical microscopy image of a typical device, along with the measurement setup and the definition of the coordinate axis, are shown in Fig.~\ref{fig:setup:spectra}(a).
The IDT pair is placed such that SAW propagates along the $x$ axis. 

To study the SAW transmission characteristics, a vector network analyzer (VNA) is connected to the IDT pair; each of the pair is labeled IDT1 and IDT2.
The SAW transmission coefficient $S_{21}$ is measured using the VNA.
In the measurements, an RF electrical signal with power $P$ and frequency $f$ is applied to IDT1 and the transmitted signal is measured at IDT2.
$P$ is fixed to 0 dBm.
First, the SAW transmittance spectrum ($|S_{21}|^2$ vs $f$) is measured to determine the SAW resonance frequency $f_\mathrm{SAW}$: the results are shown in Fig.~\ref{fig:setup:spectra}(b).
Here we apply an in-plane magnetic field $H$, whose magnitude is large enough such that contribution from magnon excitation on the $S_{21}$ signal is suppressed.
In the spectrum, the frequency at which $|S_{21}|^2$ takes a maximum is defined as $f_\mathrm{SAW} $.
$f_\mathrm{SAW}$ depends on the width ($w$) and spacing ($d$) of the IDT fingers: we find $f_\mathrm{SAW} \sim 2.375$ GHz for the fifth harmonic mode of an IDT with $w = d \sim 2~\mathrm{\mu m}$.
The relation is consistent with the sound velocity of the Y+128$^\mathrm{o}$-cut LiNbO$_{3}$ substrate.

The $\mu_0 H$ dependence of the normalized $|S_{21}|^2$ is shown by the black circles in Fig.~\ref{fig:setup:spectra}(c).
We normalize $|S_{21}|^2$ measured at each $\mu_0 H$ with that obtained at $\mu_0 H \sim -100$ mT. 
$f$ is fixed to $f_\mathrm{SAW}$.
An in-plane magnetic field with $\varphi_H=135\ \mathrm{deg}$ is applied during the $S_{21}$ measurements.
The field angle is chosen such that only the acoustic magnons couple to the SAW phonons\cite{asano2023prb}.
See Supplemental Material for the $\mu_0 H$ dependence of $|S_{21}|^2$ for the other field angles\cite{SM}.
As predicted by the model calculations, we find three minimums in the spectrum, two at $\mu_0 H \sim \pm25$ mT and one near zero field.
The calculated $|S_{21}|^2$ spectrum is shown by the blue line in Fig.~\ref{fig:setup:spectra}(c), where we now include contributions from the dipole interaction and the magnetic anisotropy fields.
(Due to these fields, the positions of the minimum shifted from $\sim \pm31$ mT.)
Good agreement is found between the experiments and the calculations.
Importantly, the minimum at zero field is clearly reproduced by the calculations, indicating the presence of the $\Phi = - \pi/2$ coupling.
Note that a previous study reported that in SAFs made of CoFeB, a material known to possess a small $\alpha$, the $|S_{21}|^2$ spectrum does not show such a minimum at zero field\cite{matsumoto2024nanolett}, corroborating the model calculations that large magnetic damping is required to observe the $\Phi = - \pi/2$ coupling.

We note that magnons near zero field no longer form the fully damped mode ($\omega_m = 0$) when the dipolar interaction and the magnetic anisotropy fields are included in the LLG equation.
Instead they fall into the category of overdamped modes, which by definition satisfy $\omega_m < \kappa_m/2$.
The emergence of the $\Phi = - \pi/2$ coupling is mostly linked with the appearance of the overdamped magnons.
See the Appendix for a more detailed discussion on this issue.

\section{CONCLUSION}
In summary, we have investigated the coupling between acoustic magnons and surface acoustic wave (SAW) phonons in a synthetic antiferromagnet (SAF) composed of Ni/Ru/Ni. 
The character of the magnon-phonon coupling is studied using the SAW transmittance. In addition to the transmittance minimums caused by the close-resonances of acoustic magnons and the SAW phonons, we find a broad local minimum near zero field in the spectrum. The latter appears in a field range where the magnon resonance frequency is far from that of the SAW and is close to zero, forming the so-called overdamped magnons. These results can be accounted for if we consider the magnon-phonon coupling constant to be a complex number. The real part describes the conventional coherent magnon-phonon coupling that occurs near the close-resonances, whereas the imaginary part is responsible for the coupling of phonons with the overdamped magnons. 
The boundary at which the coupling constant changes from real to imaginary is equivalent to an exceptional point under certain material parameters.
As the imaginary part of the coupling constant and the field range in which the overdamped regime exists scale with the magnetic damping constant, the large damping of Ni enables the observation of the imaginary coupling.
These results show that the phase of the complex magnon-phonon coupling can be tuned via magnetic field, with the imaginary coupling constant providing different means to couple magnons and SAW phonons.

\section*{Acknowledgments}
This work was partly supported by JSPS KAKENHI (Grant No. 23H05463) from JSPS and MEXT Initiative to Establish Next-generation Novel Integrated Circuits Centers (X-NICS). Y.T. acknowledges financial support from SPRING GX program from the University of Tokyo.

\appendix
\section{LANDAU LIFSHITZ GILBERT EQUATIONS}
The Landau-Lifshitz-Gilbert (LLG) equation reads
\begin{equation}
\begin{aligned}
    \dot{\bm{m}}_i=&-\gamma\bm{m}_i\times\mu_0\bm{H}_{\mathrm{eff},i}+\alpha\bm{m}_i\times\dot{\bm{m}}_i,\\
    &\mu_0\bm{H}_{\mathrm{eff},i}=-\dfrac{1}{M_{\mathrm{s}}}\dfrac{\partial E_{\mathrm{tot}}}{\partial\bm{m}_i},\\
    &E_{\mathrm{tot}}=E_{\mathrm{ext}}+E_{\mathrm{ani}}+E_{\mathrm{dem}}+E_{\mathrm{iec}}+E_{\mathrm{sd},i}+E_{\mathrm{md},i}+E_{\mathrm{me}},\\
    &\ \ \ \ \ E_{\mathrm{ext}}=-\mu_0M_{\mathrm{s}}\bm{H}\cdot(\bm{m}_1+\bm{m}_2),\\
    &\ \ \ \ \ E_{\mathrm{ani}}=-\mu_0M_{\mathrm{s}}H_{\mathrm{u}}[(\bm{m}_1\cdot\bm{e}_{\mathrm{u}})^2+(\bm{m}_2\cdot\bm{e}_{\mathrm{u}})^2],\\
    &\ \ \ \ \ E_{\mathrm{dem}}=\mu_0\dfrac{M_{\mathrm{s}}^2}{2}(m_{1z}^2+m_{2z}^2),\\
    &\ \ \ \ \ E_{\mathrm{iec}}=\mu_0M_{\mathrm{s}}H_{\mathrm{E}}(\bm{m}_1\cdot\bm{m}_2),\\
    &\ \ \ \ \ E_{\mathrm{me}}=M_{\mathrm{s}}\displaystyle\sum_{i=1,2}[b_1 \epsilon_{xx}m_{ix}^2+2b_2\epsilon_{xz}m_{ix}m_{iz}],
\end{aligned}
\end{equation}
$\bm{m}_i$ is an unit vector that represents the magnetization direction of ferromagnetic layer $i$ of the SAF. 
$M_\mathrm{s}$ is the saturation magnetization and $\alpha$ is the magnetic damping constant of the two layers.
$E_\mathrm{ext}$, $E_\mathrm{ani}$, $E_\mathrm{dem}$, $E_\mathrm{iec}$, and $E_\mathrm{me}$ are the Zeeman energy, uniaxial magnetic anisotropy energy, demagnetization energy, interlayer exchange coupling energy and magnetoelastic energy, respectively.
$E_{\mathrm{sd},i}$ and $E_{\mathrm{md},i}$ are the self- and mutual-dipole energies\cite{nortemann1993prb,stamps1994prb,shiota2020prl,asano2023prb}.
$\bm{H}$ is the external magnetic field, $\bm{H}_\mathrm{u} = H_\mathrm{u} \bm{e}_\mathrm{u}$ is the uniaxial magnetic anisotropy field, $H_\mathrm{E}$ is the interlayer exchange coupling field and $b_1$ and $b_2$ are the magnetoelastic constants.
For $E_\mathrm{me}$, we consider contributions from the $xx$ ($\epsilon_{xx}$) and $xz$ ($\epsilon_{xz}$) components of the strain tensor.

We define a $x'y'z'$ coordinate system in which the magnetization equilibrium direction points along the $z'$ axis. Using a coordinate transformation that converts from the $xyz$ laboratory frame to the $x'y'z'$ frame, the linearized LLG equation reads
\begin{equation}
\begin{aligned}
    \label{LLG_prime}
    \partial_t\bm{m}'&=A_m\bm{m}'+\bm{b}_m,\\
    &\bm{m}'=(m_{1x}',m_{1y}',m_{2x}',m_{2y}')^\top,\\
    &A_m=-\mu_0\gamma D^{-1}A,\\
    &\bm{b}_m=-\gamma D^{-1}\bm{B}.
\end{aligned}
\end{equation}
Here, $m_{1x}'$ and $m_{1y}'$ are the $x'$ and $y'$ components of the magnetization of the first layer, and $m_{2x}'$ and $m_{2y}'$ are those of the second layer.
$D$, $A$, and $\bm{B}$ are given as
\begin{equation}
\begin{gathered}
    D=\begin{pmatrix}1&\alpha&0&0\\-\alpha&1&0&0\\0&0&1&\alpha\\0&0&-\alpha&1\end{pmatrix}, \ 
    A=\begin{pmatrix}A_{11}&A_{12}&A_{13}&A_{14}\\A_{21}&A_{22}&A_{23}&A_{24}\\A_{31}&A_{32}&A_{33}&A_{34}\\A_{41}&A_{42}&A_{43}&A_{44}\end{pmatrix}, \\ 
    \bm{B}=\begin{pmatrix}b_1\epsilon_{xx}\sin{2\varphi_1}\\-2b_2\epsilon_{xz}\cos{\varphi_1}\\b_1\epsilon_{xx}\sin{2\varphi_2}\\-2b_2\epsilon_{xz}\cos{\varphi_2}\end{pmatrix}.
\end{gathered}
\end{equation}
The components of the $A$ matrix read\cite{shiota2020prl,matsumoto2024nanolett}
\begin{widetext}
\begin{equation}
\begin{aligned}
    &A_{\mathrm{11}}=0,\\
    &A_{\mathrm{12}}=H\cos{(\varphi_H-\varphi_1)}+2H_{\mathrm{u}}\cos{2(\varphi_1-\varphi_{\mathrm{u}})}-H_{\mathrm{E}}\cos{(\varphi_1-\varphi_2)}+H_{\mathrm{D}}\sin^2{\varphi_{k1}},\\
    &A_{\mathrm{13}}=-iH_{\mathrm{D}}\sin\varphi_{k1},\\
    &A_{\mathrm{14}}=H_{\mathrm{E}}\cos{(\varphi_1-\varphi_2)}+H_{\mathrm{D}}\sin{\varphi_{k1}}\sin{\varphi_{k2}},\\
    &A_{\mathrm{21}}=-M_{\mathrm{s}}+H_{\mathrm{D}}-H\cos{(\varphi_H-\varphi_1)}-2H_{\mathrm{u}}\cos^2{(\varphi_1-\varphi_{\mathrm{u}})}+H_{\mathrm{E}}\cos{(\varphi_1-\varphi_2)},\\
    &A_{\mathrm{22}}=0,\\
    &A_{\mathrm{23}}=-H_{\mathrm{E}}+H_{\mathrm{D}},\\
    &A_{\mathrm{24}}=iH_{\mathrm{D}}\sin{\varphi_{k2}},\\
    &A_{\mathrm{31}}=iH_{\mathrm{D}}\sin{\varphi_{k2}},\\
    &A_{\mathrm{32}}=H_{\mathrm{E}}\cos{(\varphi_1-\varphi_2)}+H_{\mathrm{D}}\sin{\varphi_{k2}}\sin{\varphi_{k1}},\\
    &A_{\mathrm{33}}=0,\\
    &A_{\mathrm{34}}=H\cos{(\varphi_H-\varphi_2)}+2H_{\mathrm{u}}\cos{2(\varphi_2-\varphi_{\mathrm{u}})}-H_{\mathrm{E}}\cos{(\varphi_1-\varphi_2)}+H_{\mathrm{D}}\sin^2{\varphi_{k2}},\\
    &A_{\mathrm{41}}=-H_{\mathrm{E}}+H_{\mathrm{D}},\\
    &A_{\mathrm{42}}=-iH_{\mathrm{D}}\sin{\varphi_{k1}},\\
    &A_{\mathrm{43}}=-M_{\mathrm{s}}+H_{\mathrm{D}}-H\cos{(\varphi_H-\varphi_2)}-2H_{\mathrm{u}}\cos^2{(\varphi_2-\varphi_{\mathrm{u}})}+H_{\mathrm{E}}\cos{(\varphi_1-\varphi_2)},\\
    &A_{\mathrm{44}}=0,
\end{aligned}
\end{equation}
\end{widetext}
where $\varphi_i~(i=1, 2)$ is the angle between the magnetization of layer $i$ and $x$, $\varphi_{ki}$ is the angle between the magnetization of layer $i$ and the magnon propagation direction (i.e., along $x$), and $\varphi_u$ is the angle between the magnetic easy axis of the two layers and $x$. $H_{\mathrm{D}}$ is the dipolar interaction field.
We define the eigenvalue and the corresponding eigenstates of $A_m$ as $e_{\mu}$ and $\bm{v}_{\mu}$ ($\mu = 1,2,3,4$), respectively. $e_{\mu}$ and $\bm{v}_{\mu}$ satisfy the following relation:
\begin{equation}
\begin{aligned}
    P^{-1}A_mP&=\mathrm{diag}[e_1,e_2,e_3,e_4]\equiv V,\\
    &P\equiv\begin{pmatrix}\bm{v}_1,\bm{v}_2,\bm{v}_3,\bm{v}_4\end{pmatrix}.
\end{aligned}
\end{equation}
Setting $\tilde{\bm{m}}\equiv P^{-1}\bm{m}'$, Eq.~(\ref{LLG_prime}) is expressed as
\begin{equation}
\begin{aligned}
    \label{LLG_tilde}
    &\dot{\tilde{\bm{m}}}=V\tilde{\bm{m}}+P^{-1}\bm{b}_m,\\
    &\tilde{\bm{m}}=(\tilde{m}_1,\tilde{m}_2,\tilde{m}_3,\tilde{m}_4)^\top.
\end{aligned}
\end{equation}
The four states with $\mu=1,2,3,4$ represent the optical and acoustic magnon modes with positive and negative eigenfrequencies.

\section{THE COUPLED EQUATIONS OF MOTION}
\subsection{Equations of motion}
To simplify discussion, we consider one magnon eigenmode among the four that derives from Eq.~(\ref{LLG_tilde}).
We pick one state $\tilde{\bm{m}}_\mu$ with eigenvalue $e_\mu$ and express them as $M$ and $e$, respectively~\cite{asano2023prb}.
Following the discussion in Ref.~\cite{asano2023prb} and Eq.~(\ref{LLG_tilde}), the elastic wave equation and the LLG equation read 
\begin{align}
    \label{acoustic wave eq}
    &\ddot{U}+\kappa_a\dot{U}+\omega_a^2U=-2\omega_a g_{am}M,\\
    \label{LLG eq}
    &\dot{M}=e M-ig_{ma}U=-( \frac{\kappa_m}{2} + i\omega_m ) M-ig_{ma}U
\end{align}
where $U$ ($M$), $\kappa_a/2$ ($\kappa_m/2$), $\omega_a$ ($\omega_m$) are the amplitude, relaxation rate and the frequency of the SAW phonons (magnons). 
$g_{am}$ and $g_{ma}$ are the coupling constants that originate from the magnetoelastic coupling energy $E_\mathrm{me}$.

Setting $U=ae^{i(kx-\omega t)}$ and $M=me^{i(kx-\omega t)}$ in Eqs.~(\ref{acoustic wave eq}) and (\ref{LLG eq}), where $k$ is the wave vector of the SAW phonon defined by the pitch of the IDT fingers and $\omega$ is the frequency of the rf signal applied to the IDT, we have
\begin{align}
    &\ddot{a}+(\kappa_a-2i\omega)\dot{a}+(\omega_a^2-\omega^2-i\omega\kappa_a)a=-2\omega_ag_{am}m,\\
    &\dot{m}-i\omega m=-\dfrac{\kappa_m}{2}m-i\omega_mm-ig_{ma}a.
\end{align}
Here, we ignore propagation decay of $U$ along the piezoelectric substrate and thus set $k$ as a real number.
Employing the rotating wave approximation, one obtains
\begin{align}
    \label{acoustic}
    &\dot{a}=\left[i(\omega-\omega_a)-\dfrac{\kappa_a}{2}\right]a-ig_{am}m,\\
    \label{LLG}
    &\dot{m}=\left[i(\omega-\omega_m)-\dfrac{\kappa_m}{2}\right]m-ig_{ma}a.
\end{align}
Equations~(\ref{acoustic}) and (\ref{LLG}) are equivalent to Eq.~(1) in the main text.

\subsection{The coupling constant}
Following Ref.~\cite{asano2023prb}, the coupling constant $g$ and its phase $\Phi$ are given as the following. 
Here we show the coupling constant of all four magnon modes.
\begin{widetext}
\begin{equation}
\begin{aligned}
    g_{\mu}e^{i\frac{\Phi_{\mu}}{2}}&=\sqrt{g_{am,\mu}g_{ma,\mu}},\\
    \begin{pmatrix}g_{am,1} & g_{am,2} & g_{am,3} & g_{am,4}\end{pmatrix}&=-\dfrac{ikM_{\mathrm{s}}S_m}{4\omega_a\rho S_a}\begin{pmatrix}2\eta_zb_2\cos{\varphi_1} & b_1\sin{2\varphi_1} & 2\eta_zb_2\cos{\varphi_2} & b_1\sin{2\varphi_2}\end{pmatrix}P\\    \begin{pmatrix}g_{ma,1}\\g_{ma,2}\\g_{ma,3}\\g_{ma,4}\end{pmatrix}&=\dfrac{k\gamma}{1+\alpha^2}P^{-1}\begin{pmatrix}b_1\sin{2\varphi_1}+2\eta_zb_2\alpha\cos{\varphi_1}\\b_1\alpha\sin{2\varphi_1-2\eta_zb_2\cos{\varphi_1}}\\b_1\sin{2\varphi_2}+2\eta_zb_2\alpha\cos{\varphi_2}\\b_1\alpha\sin{2\varphi_2-2\eta_zb_2\cos{\varphi_2}}\end{pmatrix}.
    \label{eq:coupling}
\end{aligned}
\end{equation}
\end{widetext}
$\rho$ is the density of the media, $\eta_z\equiv\epsilon_{xz}/\epsilon_{xx}$ is the ratio of the $xz$ and $xx$ components of the strain and $S_m$ and $S_a$ are the effective area of the magnons and phonons.

\subsection{SAW transmission}
To derive the SAW transmittance $T$, which is equivalent to $|S_{21}|^2$ obtained in the experiments, we set $U(x,t)=ae^{-\frac{r}{2}x+i(kx-\omega t)}$, $M(x,t)=me^{i(kx-\omega t)}$, where Re[$r$] represents the reduction of SAW amplitude due to magnon-phonon coupling. Neglecting contributions of $\mathcal{O}(r^2)$, we obtain
\begin{align}
    \label{deltakappa_a}
    \mathrm{Re}[r]=\dfrac{\delta\kappa_a}{v},~\delta\kappa_a=-2\sum_\mu\mathrm{Im}\left[\dfrac{i\left(g_\mu e^{i\frac{\Phi_\mu}{2}}\right)^2}{i\omega_a+e_\mu}\right]
\end{align}
$\delta\kappa_a$ represents coupling-induced modulation of phonon relaxation\cite{asano2023prb}.
The normalized SAW phonon transmittance $T$ is obtained from the following definition:
\begin{align}
    \label{transmittance}
    T=e^{-\mathrm{Re}[r]L_x}=e^{-\frac{\delta\kappa_aL_x}{v}}
\end{align}
$L_x$ is the length of the SAF along the SAW propagation direction.

\section{EFFECTS OF DIPOLAR INTERACTION AND ANISOTROPY FIELDS}
\begin{figure}[b]
	\centering
	\includegraphics[width=1\linewidth]{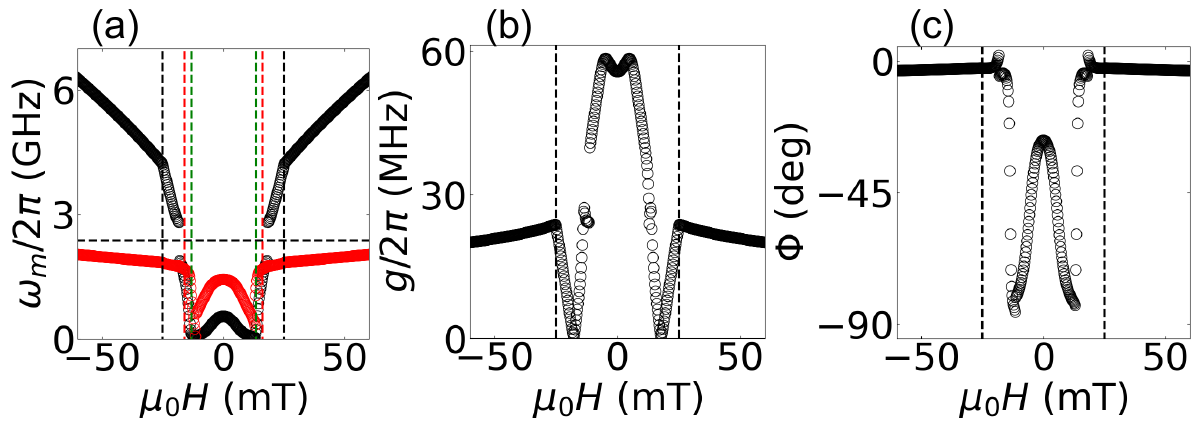}
	\caption{(a)-(c) Magnetic field $\mu_0 H$ dependence of the acoustic magnon eigenfrequency $\omega_m / 2 \pi$ (black circles) and the relaxation rate $(\kappa_m/2)/2\pi$ (red circles) (a), the magnitude $g$ (b) and the phase $\Phi$ (c) of the magnon-phonon coupling constant. The vertical red and green dashed lines in (a) indicate the boundary between the overdamped and underdamped regimes, and $|H| = H_\mathrm{tp}$, respectively. The horizontal dashed line in (a) shows the SAW resonance frequency used in the experiments. The vertical black dashed lines in (a-c) indicate $\mu_0H = \pm25$ mT. Parameters used in the calculations are listed in Table~\ref{table:par}. Here the uniaxial magnetic anisotropy field $\mu_0 H_\mathrm{u}$ and the dipolar interaction field $\mu_0 H_\mathrm{D}$ are included.
    \label{fig:coupling_sup}
	}
\end{figure}

In the discussion pertaining to Fig.~\ref{fig:transmission} of the main text, we neglected the dipolar interaction and magnetic anisotropy fields. Here we show how these fields influence the coupling constant and the SAW transmittance. 
The black circles in Fig.~\ref{fig:coupling_sup}(a) show the $\mu_0 H$ dependence of the eigenfrequency of the acoustic-mode magnons in the symmetric SAF.
The corresponding magnitude $g$ and phase $\Phi$ of the magnon-phonon coupling constant are shown in Figs.~\ref{fig:coupling_sup}(b) and \ref{fig:coupling_sup}(c), respectively.
The parameters used are the same with those of Fig.~\ref{fig:setup:spectra}(c) of the main text.
As is evident, inclusion of the dipolar interaction and magnetic anisotropy fields causes the magnon eigenfrequency $\omega_m / 2 \pi$ near zero field to take nonzero values.
Thus the magnons in such field range no longer belong to the fully damped mode but are categorized as overdamped magnons, which satisfy $\kappa_m/2 > \omega_m$.
Note that we cannot derive an analytical formula for $H_\mathrm{tp}$ [as in Eq.~(\ref{eq:hep}) of the main text] when the magnetic anisotropy and dipolar interaction fields are included.
We, therefore, simply denote the field at which the coupling changes from $\Phi = 0$ to $-\pi/2$ as $H_\mathrm{tp}$ hereafter.

The horizontal dashed line in Fig.~\ref{fig:coupling_sup}(a) represents the SAW resonance frequency.
Due to the dipolar interaction and magnetic anisotropy fields, a gap opens near the SAW resonance frequency. 
One may thus assume that the acoustic magnon-SAW coupling is negligible. 
Owing to the large magnetic damping, however, we show that the coupling is nonzero and causes minimums in the SAW transmittance spectrum.

\begin{figure}[b]
	\centering
	\includegraphics[width=1.0\linewidth]{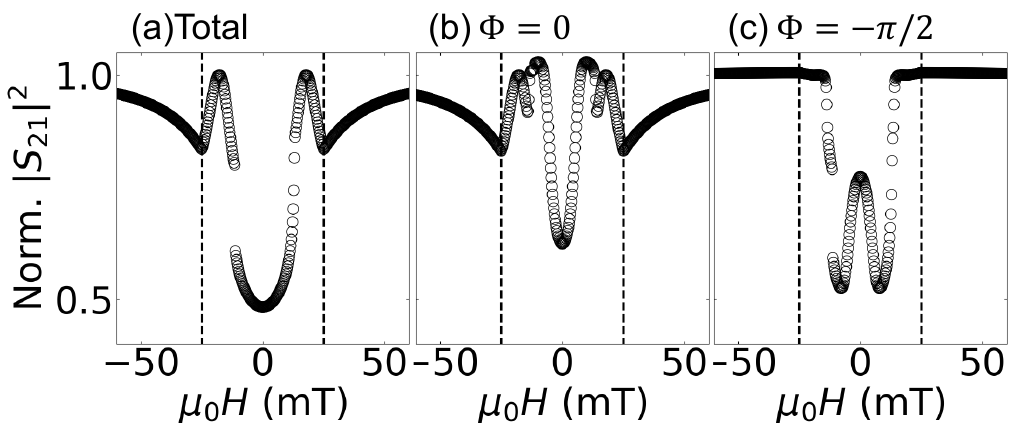}
	\caption{
    (a)-(c) Normalized SAW transmittance $|S_{21}|^2$ vs in-plane magnetic field $\mu_0 H$. $|S_{21}|^2$ is calculated assuming $\Phi = 0$ (b) and $\Phi = -\pi/2$ (c). The sum of the two is presented in (a). The vertical dashed lines indicate $\mu_0H = \pm25$ mT. Parameters used in the calculations are listed in Table~\ref{table:par}. Here the uniaxial magnetic anisotropy field $\mu_0 H_\mathrm{u}$ and the dipolar interaction field $\mu_0 H_\mathrm{D}$ are included.  
	\label{fig:transmission_ani}
	}
\end{figure}

Figures~\ref{fig:transmission_ani}(a)-\ref{fig:transmission_ani}(c) show the in-plane external magnetic field $\mu_0H$ dependence of $|S_{21}|^2$ when the dipolar interaction and anisotropy fields are included.
The plot in Fig.~\ref{fig:transmission_ani}(a) is the same as that presented by the blue solid line in Fig.~\ref{fig:setup:spectra}(c) of the main text.
Contributions from $\Phi = 0$ and $-\pi/2$ couplings are shown in Figs.~\ref{fig:transmission_ani}(b) and \ref{fig:transmission_ani}(c).
As is evident, the transmission dip at $\mu_0 H\sim\pm25$ mT is caused by the $\Phi=0$ coupling, whereas the dip near zero field is primarily due to the $\Phi = -\pi/2$ coupling.
In contrast to the case without the dipolar interaction and magnetic anisotropy fields [Figs.~\ref{fig:coupling}(b)-\ref{fig:coupling}(d)], here the $\Phi=0$ coupling also contributes to the dip near zero field.
This goes against the notion that transmittance dip due to coherent coupling ($\Phi=0$) must occur when the magnon and phonon resonance conditions match. 
In the following, we describe the reasoning behind.

First, we discuss the $\Phi = 0$ coupling.
With $\Phi = 0$, the imaginary part of $(ge^{i\frac{\Phi}{2}})^2$ becomes zero: see Eq.~(\ref{eq:geff}) of the main text.
Under such condition, Eq.~(\ref{deltakappa_a}) can be rewritten as
\begin{align}
    \label{delta_kappa_a_rev}
    \delta\kappa_a&=-2\mathrm{Im}\left.\left[\dfrac{i\left(ge^{i\frac{\Phi}{2}}\right)^2}{i\omega_a+e}\right]\right|_{\Phi=0}\nonumber\\
    &=2g^2\times\dfrac{\kappa_m/2}{(\omega_m-\omega_a)^2+(\kappa_m/2)^2}.
\end{align}
$\delta\kappa_a$, which represents the SAW transmittance [Eq.~(\ref{transmittance})], is equal to the product of the coupling constant square ($g^2$) and a Lorentzian function that describes the resonance of magnons with phonons. 
Note that there is a one-to-one correspondence between $\omega_m$ and $\mu_0 H$ when $|H| > H_\mathrm{tp}$: here $\omega_m$ scales with $\mu_0 |H|$.
($g^2$ is also a function of $\mu_0 H$: we discuss its effect in the next paragraph.)
Thus Eq.~(\ref{delta_kappa_a_rev}) can be interpreted as the $\mu_0 H$ dependence of $\delta\kappa_a$.
The Lorentzian function peaks at the resonance condition, $\omega_m = \omega_a$.
As the SAW transmittance drops at a field where $\delta \kappa_a$ takes a maximum, we expect the transmittance to exhibit a dip when the resonance condition is satisfied.
This is the generally accepted origin of the SAW transmittance dip due to the coherent ($\Phi = 0$) coupling.

The transmittance dip can also occur when the $\mu_0 H$ dependence of $g^2$ (or $g$) exhibits sharp features while the Lorentzian function takes a nonzero value.
The amplitude of the Lorentzian function is generally small (close to zero) outside its peak structure.
However, when $\kappa_m$ is large, i.e., in materials with large magnetic damping, it takes nonzero values that allow structures of $g^2$ to show up in the transmittance spectrum.
From the field dependence of $g$ presented in Fig.~\ref{fig:coupling_sup}(b), one finds that $g$ tends to increase with decreasing field.
This is due to the $\mu_0 H$ dependence of the magnetoelastic coupling, which in most cases scales with $1/|H|$. 
In addition, for SAF, $g$ shows significant variation with $\mu_0 H$: it is zero at $|H| = \sqrt{2} H_\mathrm{E}$ and takes a local maximum at $|H| = 2 H_\mathrm{E}$.
Comparing the results presented in Figs.~\ref{fig:transmission_ani} and \ref{fig:coupling_sup}, we find the transmittance dips found near $\mu_0 H \sim \pm25$ mT coincide with the peak in $g/2\pi$: see the vertical dashed lines in Figs.~\ref{fig:transmission_ani}(b) and Fig.~\ref{fig:coupling_sup}(b).
Moreover, owing to the enhancement of $g$ at $\mu_0 H \sim 0$, the $\Phi = 0$ coupling contributes to the SAW transmittance minimum near zero field.
These results thus demonstrate that the $\mu_0 H$ dependence of $g^2$ in Eq.~(\ref{delta_kappa_a_rev}) can influence the SAW transmittance when the magnetic damping is large.
Note that here a uniform magnetized state is assumed throughout the calculations.
If a multidomain state was to appear, e.g., at around zero field, the size of the dip will simply reduce. 
Thus formation of such multidomain state cannot account for the experimental results.
(We neglect interaction of SAW with domain walls\cite{zhang2026prb} due to large SAW wavelength compared to the size of domain walls.) 

Next, we discuss the $\Phi = -\pi/2$ coupling.
In this case, $(ge^{i\frac{\Phi}{2}})^2$ is a pure imaginary number: see Eq.~(\ref{eq:geff}) of the main text for the definition of the coupling constant.
With $\Phi = -\pi/2$, Eq.~(\ref{deltakappa_a}) is re-expressed as
\begin{align}
    \label{delta_kappa_a_rev:diss}
    \delta\kappa_a&=-2\mathrm{Im}\left.\left[\dfrac{i\left(ge^{i\frac{\Phi}{2}}\right)^2}{i\omega_a+e}\right]\right|_{\Phi=-\frac{\pi}{2}}\nonumber\\
    &=-2g^2\times\dfrac{\omega_m-\omega_a}{(\omega_m-\omega_a)^2+(\kappa_m/2)^2}.
\end{align}
As is evident, $\delta\kappa_a$ is zero under the resonance conditions of magnons and phonons, i.e., when $\omega_m = \omega_a$.
Thus, the $\Phi = -\pi/2$ coupling does not cause a SAW transmission dip when $\omega_m = \omega_a$. 
Rather, $\delta\kappa_a$ reflects the field dependence of $g^2$ when $\omega_m$ and $\omega_a$ are different.
Since the imaginary part of $(ge^{i\frac{\Phi}{2}})^2$ is negligible when $|H| > H_\mathrm{tp}$ [see Figs.~\ref{fig:coupling_sup}(b) and \ref{fig:coupling_sup}(c)], the coupling manifests itself when $|H| \leq H_\mathrm{tp}$.

Finally, we show in Fig.~\ref{fig:coupling_sup}(a), the boundary that separates the overdamped ($\kappa_m/2 > \omega_m$) and underdamped ($\kappa_m/2 < \omega_m$) regimes. 
We show the value of $(\kappa_m/2)/2\pi$ in Fig.~\ref{fig:coupling_sup}(a) using the red dotted line. The boundary of the two regimes, i.e., the field at which $\kappa_m/2 = \omega_m$, is shown using the vertical red dashed line in Fig.~\ref{fig:coupling_sup}(a). In the parameter set used here, the boundary is close to $H_\mathrm{tp}$, represented by the vertical green dashed line. We note that the boundary is always larger than $H_\mathrm{tp}$, indicating that the magnon modes that contribute to the $\Phi = -\pi/2$ coupling are always the overdamped magnons.

\section{EIGENENERGY OF THE MAGNON MODES AND THE EXCEPTIONAL POINT}
\begin{figure}[t]
	\centering
	\includegraphics[width=0.8\linewidth]{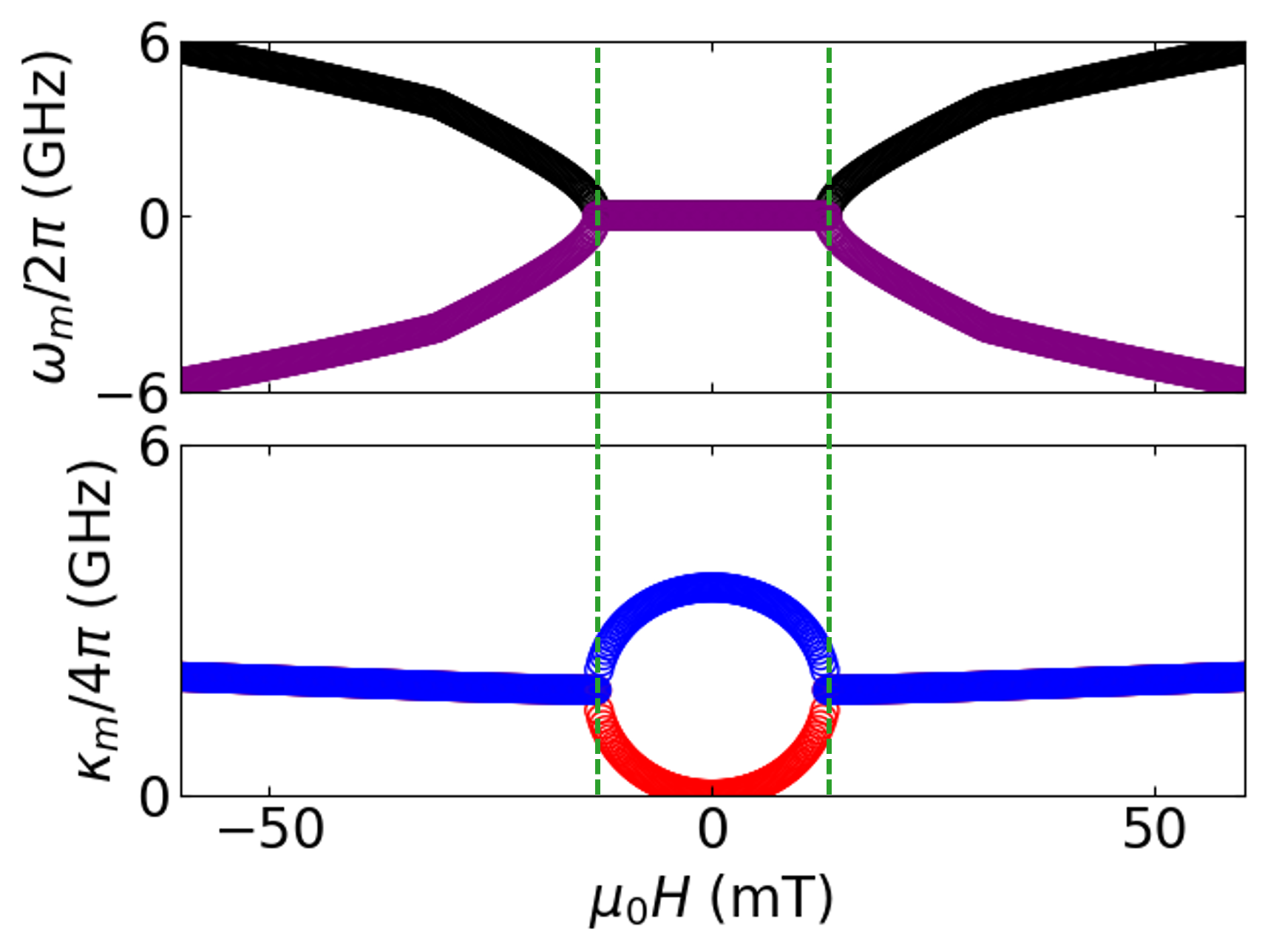}
	\caption{Magnetic field $\mu_0H$ dependence of the acoustic magnon eigenfrequency $\omega_m/2\pi$ (top) and the relaxation rate $\kappa_m/4\pi$ (bottom). The black and red circles correspond to the positive-frequency mode, whereas the purple and blue circles represent the negative-frequency mode. The vertical green dotted lines indicate the transition point $H_{\mathrm{tp}}$. The uniaxial magnetic anisotropy field $\mu_0 H_\mathrm{u}$ and the dipolar interaction field $\mu_0 H_\mathrm{D}$ are set to zero.
    \label{fig:Negative frequency mode magnon}
	}
\end{figure}

Here we show the evolution of eigenenergy of the SAF with the magnetic field to identify the appearance of the exceptional point. 
The eigenenergy has real and imaginary parts, which correspond to the relaxation rate $\kappa_m / 4 \pi$ and the eigenfrequency $\omega_m / 2 \pi$, respectively. 
The magnetic field dependence of $\omega_m / 2 \pi$ and $\kappa_m / 4 \pi$ of the acoustic magnon modes are presented in top and bottom panels of Fig.~\ref{fig:Negative frequency mode magnon}, respectively. 
Here, the magnetic anisotropy and dipolar interaction fields are neglected in order to provide comparison with the analytical model described in the main text.
As is evident from top panel of Fig.~\ref{fig:Negative frequency mode magnon}, the two acoustic modes have positive and negative eigenfrequencies.
The black and red circles in Fig.~\ref{fig:Negative frequency mode magnon} correspond to the positive-frequency mode, whereas the purple and blue circles represent the negative-frequency mode. The green dotted line indicates the transition point $H_\mathrm{tp}$.
At $|H| \leq H_\mathrm{tp}$, the eigenfrequencies of the two modes coalesce and become zero. The relaxation rate, in contrast, is identical for the two acoustic modes when $|H| \geq H_\mathrm{tp}$, however, they diverge when $|H| \leq H_\mathrm{tp}$. From the plot, one can identify that the exceptional points emerge at $|H| = H_\mathrm{tp}$.

We emphasize that the exceptional point discussed here should be understood as a mathematical property of the non-Hermitian dynamical matrix [$V$ in Eq.~(\ref{LLG_tilde})] governing the magnetization dynamics, rather than the coalescence of two independent physical modes observed in other systems\cite{zhang2017ncomm}. While the positive and negative frequency branches are not experimentally accessible in the same manner as conventional two-mode systems, the eigenfrequency branching and the underdamped-to-overdamped transition exhibit characteristics analogous to those of conventional exceptional points.

\subsection*{Parameters used in the calculations}
Parameters used in the calculations are presented in Table~\ref{table:par}.

\begin{widetext}
\begin{center}
\begin{table}[h]
  \caption{Definition and value of the parameters used in the numerical calculations.}
  \label{table:par}
  \centering
  \begin{tabular}{cccc}
    \hline
    Symbol & Parameter & Value & Source \\
    \hline \hline
    $v$ & Velocity of LiNbO3 & $3950~\mathrm{m~s^{-1}}$ & Exp.\\
    $k$ & SAW wave number & $3.8\times10^6~\mathrm{m}^{-1}$ & Designed\\
    $t_{\mathrm{F}}$ & Thickness of the ferromagnetic layer & $5~\mathrm{nm}$ & Designed\\
    $b_1$ & Magnetoelastic constant (longitudinal strain) & $14~\mathrm{T}$ & Ref.~\cite{hatanaka2022prap}\\
    $\rho$ & Mass density\footnote{$\rho$ of  $\mathrm{LiNbO_3}$ is used.} & $4650~\mathrm{kg~m^{-3}}$ & Ref.~\cite{kushibiki1999ieee}\\
    $\eta_z$ & Ratio of strains $\epsilon_{xz}$ and $\epsilon_{xx}$ & $0$ & Ref.\cite{kawada2025jap}\\
    $\alpha$ & magnetic damping constant & $0.185$ & Fitting\\
    $M_{\mathrm{s}}$ & Saturation magnetization & $4.91\times 10^{5}~\mathrm{A~m^{-1}}$ & Ref.~\cite{crangle1971roysoclond}\\
    $\gamma$ & Gyromagnetic ratio & $1.92\times 10^{11}~\mathrm{T^{-1}s^{-1}}$ & Ref.\cite{hatanaka2022prap}\\
    $\mu_0H_{\mathrm{E}}$ & Interlayer exchange coupling field  & $15.5~\mathrm{mT}$ & Exp.\footnote{$\mu_0H_\mathrm{E}$ are obtained using transport measurements\cite{matsumoto2022apex, SM}.}\\
    $\mu_0H_{\mathrm{u}}$ & Uniaxial magnetic anisotropy field & $3.0~\mathrm{mT}$ & Fitting\\
    $\varphi_\mathrm{u}$ & Direction of $H_\mathrm{u}$ within the film plane  & $130~\mathrm{deg}$ & Fitting\\
    $S_m/S_a$ & Ratio of the magnon phonon effective area & $5.4\times10^{-3}$ & Designed\footnote{Ratio of the area of the thin film and the SAW propagation line.}\\
    $\mu_0H_{\mathrm{D}}$ & Magnetic dipole field & $5.9~\mathrm{mT}$ & Exp.\footnote{Calculated using the relation: $H_{\mathrm{D}}=M_{\mathrm{s}}(1-e^{-2kt})/4$\cite{shiota2020prl}.}\\
    $L_x$ & Length of SAF along SAW & $450~\mathrm{\mu m}$ & Designed\\
    \hline\hline
  \end{tabular}
\end{table}
\end{center}
\end{widetext}

\clearpage
\bibliography{refs_050125}

\end{document}